# Soap Films Amplify the Air–Water Interface but Neither Produce Detectable $H_2O_2$ Nor Enhance Reduction of Gold Ions

Muzzamil Ahmad Eatoo[1,2,*], Yash K. Mohature[1,2], Francisco P. Tome[1,2], Lisa Exposito[1,2], Sirisha Kamireddy[1,2], Najeh Kharbatia[1,2], and Himanshu Mishra[1,2,*]

[1] Environmental Science and Engineering (EnSE) Program, Biological and Environmental Science and Engineering (BESE) Division, King Abdullah University of Science and Technology (KAUST), 23955-6900, Thuwal, Kingdom of Saudi Arabia

[2] Water Desalination and Reuse Platform (WDRP), Biological and Environmental Science and Engineering (BESE) Division, King Abdullah University of Science and Technology (KAUST), 23955-6900, Thuwal, Kingdom of Saudi Arabia

*Correspondence: muzzamil.eatoo@kaust.edu.sa and himanshu.mishra@kaust.edu.sa

**Abstract:**

Soap films afford gentle production of air–water interfaces and underlie numerous practical applications. Recently these interfaces have been shown to exhibit unexpected redox capability, wherein Triton X-100/PBS soap films formed with water or 100 μM $HAuCl_4$ solutions (i) oxidize water up to ~1.67 mM hydrogen peroxide ($H_2O_2$) and (ii) reduce aqueous $Au^{3+}$ ions into gold nanoparticles, without external energy or catalyst. These reports tend to align with the broader claims of the redox nature of the air-water interface, which has been intensely debated. Here, we re-examine these claims via complementary application of a broad range of techniques, including $^1$H-NMR spectroscopy, fluorescence, colorimetric, transmission electron microscopy, dynamic light scattering, and inductively coupled plasma-optical emission spectroscopy (ICP-OES). Experimental results reveal no evidence for $H_2O_2$ formation despite repeated formation of over 300 cycles of soap films, were probed via $^1$H NMR spectroscopy (detection limit of 50 nM) and other analytical methods. Next, we interrogated soap films for their ability to spontaneously reduce $Au^{3+}$ ions. TEM and time-resolved DLS revealed that the formation of Au nanoparticles was already formed spontaneous in bulk solutions, and casting soap films (over 300 cycles) had no measurable effect on particle size or density. This observation was confirmed via complementary quantitative ICP-OES. Our investigations showed that Au nanoparticles formation in bulk 100 μM $HAuCl_4.3H_2O$ solution without or with 0.1% triton soap was accompanied by $H_2O_2$ generation, and we provide mechanistic insights into these processes. Collectively, these results demonstrate that amplification of the air–water interfacial area through soap-film formation is insufficient, by itself, to induce detectable $H_2O_2$ generation or enhance Au-ion reduction. More broadly, our findings highlight the need to distinguish reactions that are driven by purely air–water interfacial effects from those that originate from bulk-solution chemistry and at the solid–water interface.

## Introduction

In the last two decades, microdroplet chemistry has emerged as a rapidly expanding area of research, with numerous studies reporting remarkably accelerated reactions and unusual chemical transformations in micron-sized aqueous droplets and related interfacial systems[1-20]. Among the various phenomena reported, the spontaneous formation of hydrogen peroxide ($H_2O_2$) in water microdroplets – pneumatically sprayed or condensed – ranks among the most widely investigated and debated case-studies in current chemical science[21-28]. Here, the word *spontaneous* carries the same meaning as defined by the authors of the original reports[20-22, 24] that showed the interfacial to be a thermodynamically downhill reaction ($\triangle_{rxn} G^o < 0$) demanding no "*chemical reagent, catalyst, applied electric potential, or radiation*"[21]. These reports have therefore stimulated considerable interest in understanding whether the air–water interface possesses chemical properties distinct from those of bulk water and, if so, what factors and molecular processes are responsible for such enhanced reactivity. Mechanisms put forth to explain this unusual observation at the air–water interface include: (i) the presence of ultrahigh electric fields that causes formation of hydroxyl radicals from interfacial hydroxide ions (Refs.[21, 22, 24, 29, 30]), (ii) the spontaneous formation of water radical cations and hydrated electrons from water molecules[31], and (iii) formation of charged droplets during spraying[32], followed by dehydration of interfacial hydronium and hydroxide ions leading to $HO^\bullet$ and $H^\bullet$ radicals forming $H_2O_2$ and $H_2$[33, 34] (also see Refs.[35-38]). Naturally, if the skin of bulk water is so reactive, its environmental and practical implications would be enormous.

The abovementioned studies' conclusions and mechanisms have been called into question. Alternative explanations have been put forth based on two-electron reduction of dissolved oxygen at solid–water interfaces[39, 40], energy deposition during droplet generation via ultrasonication[41, 42], electrical voltage[43] and hydrodynamic nebulization and evaporative concentration[44], triboelectric effects[45], and incorrect attribution of mass spectrometric peaks in prior reports[46-49]. Latest reports based on vibrational sum-frequency generation spectroscopy[50] and quantum chemical calculations[36, 51, 52] have found no evidence for the proposed buildup of static or fluctuating electric fields; as well, electrochemical studies on oil–water emulsions[53] and freezing/thawing studies of water in contact with polymers[54] have also corroborated $O_2$ reduction at solid–water interfaces and triboelectric effect as the primary contributors. Yet the debate rages (see Refs.[28, 29, 35, 55-82]).

In a recent study, Dick & co-workers[83] utilized soap films using an aqueous solution of 0.1% (v/v) of Triton X-100 surfactant in a 200 mM phosphate buffer saline (PBS) as a model system for probing the intrinsic reactivity of the air–water interface. They reasoned that if unusual redox chemistry originates from the interface itself, it should also occur in geometries with large surface-area-to-volume (*SA:V*) ratios without requiring abovementioned energetic droplet-generation processes. They interrogated spontaneous $H_2O_2$ formation in Triton X-100/PBS soap films formed using bubble wands composed of polymers (polytetrafluorothylene and polyamide-imide) and metals (Ag/AgCl, Pt, and Zn). The observed $H_2O_2$ concentrations in individual soap films made with polymeric wands ranged from 0.5–2.5 mM, while those formed with metallic wands were ~0.5 mM. The authors attributed these observations to water oxidation at the air–water interface opining, "*the formation of these high SA:V ratio geometries might be a driving force behind the unusual chemistry observed in microdroplets*". One wonders that in this system, where the bulk surfactant concentration (~1.6 mM, see SI Section S7 ) exceeds the critical micelle concentration (0.22 mM) and the air–water interface would consequently be saturated with surfactants, why would the $H_2O_2$ concentrations shoot up by a factor of 1000-× higher than that in the latest reports of Zare & co-workers[24, 28, 66].

More recently, Dick & co-workers[84] extended this concept by utilizing 0.1% (v/v) of aqueous Triton X-100 surfactant solutions to show spontaneous reduction of $Au^{3+}$ ions in 100 μM solutions of $HAuCl_4.3H_2O$ leading to spontaneous formation of gold nanoparticles. They suggested that "*the in situ-generated hydrogen peroxide can spontaneously reduce metal ions in the microfilm.*" While their experimental observations seem to corroborate the findings of Zare & co-workers[20, 85], the suggested mechanism does not. Zare & coworkers[20] had suggested that the reduction of metal ions was driven by electron transfer at the air–water interface aided by the ultrahigh electric fields[20]. Both of these reports, however, agree in the fact that Au nanoparticle formation is infeasible in the bulk solution, which is not true. We have recently demonstrated that 100 μM solutions of $HAuCl_4$(aq) experience spontaneous formation of gold nanoparticles, i.e., within the seconds of forming bulk solution, and spraying water into microdroplets is not the primary driver behind nanoparticles formation[86]. This bulk redox process, in fact, entails reduction of $Au^{3+}$ ions into Au atoms that form nanoparticles, while water stoichiometrically gets oxidized to $H_2O_2$.[86]

In light of these debates and experimental irreproducibility[77-80], we re-examine the claims of spontaneous $H_2O_2$ generation and reduction of $Au^{3+}$ ions in soap films. Specifically, we strive to address the following questions:

1. Do soap-films (as reported in ref.[83]) spontaneously generate $H_2O_2$?
    a. If yes, does increasing the air–water interfacial area through repeated soap-film formation lead to an increasing $H_2O_2$ concentration in the bulk soap solution?
2. Do soap-films (as reported in ref.[84]) spontaneously reduce $Au^{3+}$ ions to Au nanoparticles due to purely interfacial effects?
    a. Do $Au^{3+}$ ions also spontaneously get reduced in bulk aqueous solutions?

Our experimental results demonstrate that water remains chemically stable during soap-film formation, with no detectable generation or accumulation of $H_2O_2$ even after repeated film-formation cycles. We further pinpoint that although Au nanoparticles readily form under the investigated conditions, their formation originates from spontaneous redox chemistry occurring in the bulk solution rather than from the soap-film interface.

**Results:**

To investigate the reported millimolar-level formation of $H_2O_2$ in soap films, we established a simple experimental setup closely following the previously reported procedure (see Methods).[83] The soap solution was prepared by mixing 0.1% (v/v) Triton X-100 with 200 mM phosphate-buffered saline in a glass Petri dish. Soap films were generated using a commercial plastic wand, and film formation was repeated multiple times (up to 300 soap films collected), with each film returned to the same solution, allowing us to interrogate $H_2O_2$ accumulation as a function of number of soap films cycles. Multiple complementary analytical techniques, including semi-quantitative peroxide test strips (as employed in the original report), Amplex Red, a Hydrogen Peroxide Assay Kit (HPAK), potassium titanium oxalate (PTO), iodometric analysis, and high resolution $^1$H NMR spectroscopy were exploited to quantify $H_2O_2$ formation. It is worth noting that our $^1$H NMR protocol developed based on the work of Bax and co-workers[87] enabled direct detection of $H_2O_2$(aq) at a detection limit $\geq$ 50 nM (See Figure S8 for NMR calibration curve).

Whereas the original study reported a pronounced color response on the peroxide test strips following soap-film formation, we observed no comparable color development under the same experimental conditions (Figure S1 and Supplementary Movie S1). These results provided no evidence for detectable $H_2O_2$ formation associated with soap-film generation. We repeated the soap film formation events because the previous report suggested that $H_2O_2$ accumulation increases with the number of soap-film formation cycles. In contrast, even after 300 consecutive cycles of soap film formation, dispensing back into the bulk solution and reforming under our experimental conditions, the peroxide test strips showed no discernible color change, providing no evidence for detectable $H_2O_2$ accumulation. After this quick testing, we resorted to advanced spectroscopic characterization.

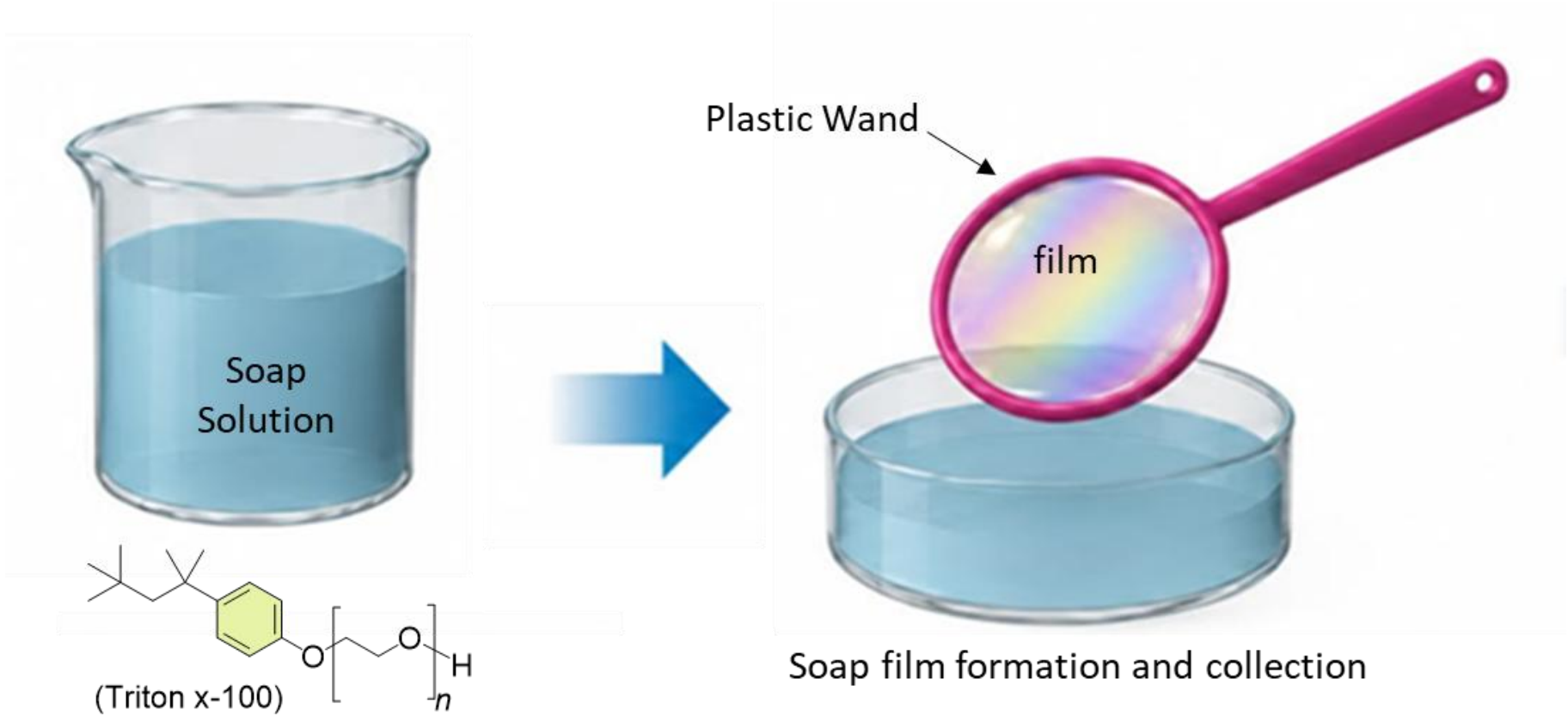

**Figure 1.** Experimental procedure for soap-film formation and sample collection. Aqueous soap solution containing 0.1% (v/v) Triton X-100 in 200 mM PBS was used to generate soap films with a commercial plastic wand. The films were formed, dispensed into the bulk, and reformed for over 300 cycles and $H_2O_2$ concentration was quantified via multiple analytical techniques.

Next, we employed $^1H$ NMR spectroscopy for $H_2O_2$ detection, following the protocol developed by Bax & coworkers which enabled a detection limit of ≥50 nM (See Figure S8). For each measurement, 480 µL of the soap films containing sample (around 200 and 300 soap film cycles sample) was added with 20 µL of $D_2O$ and transferred to a standard 5 mm quartz NMR tube. The results obtained and presented in Figure 2 revealed no detectable $H_2O_2$ formation or accumulation. To further exclude the possibility that components of the soap matrix suppressed or obscured the $H_2O_2$ peak in $^1H$ NMR analysis, we performed a spike-recovery experiment. The sample collected after 300 soap-film formation cycles was spiked with 1 µM $H_2O_2$ and reanalyzed under identical NMR conditions. A clear $H_2O_2$ peak was recovered in the spiked sample (Figure 2b), demonstrating that $H_2O_2$ remains readily detectable in the soap films matrix. Thus, the absence of an $H_2O_2$ signal after repeated soap-film formation cycles cannot be attributed to matrix-induced suppression of its NMR detection and instead indicates that no detectable $H_2O_2$ accumulated under these conditions.

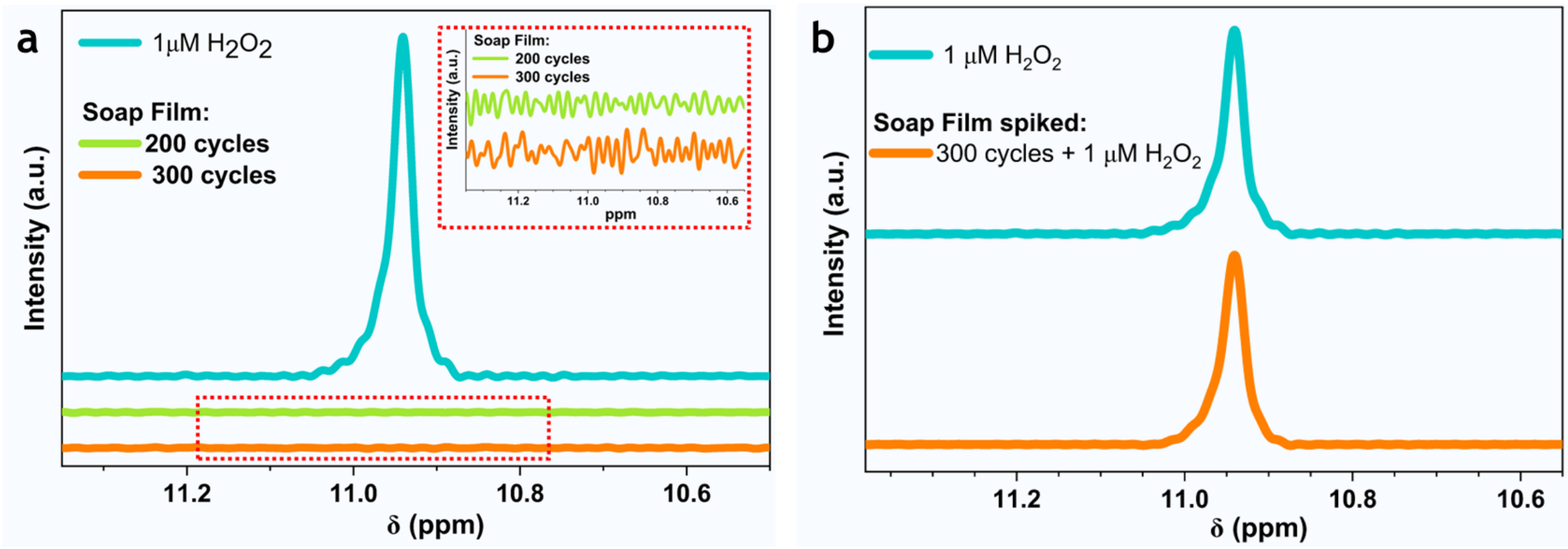

**Figure 2.** [1]H-NMR analysis of $H_2O_2$ following repeated soap-film formation and spike-recovery control. **(a)** [1]H NMR spectra of a 1 μM $H_2O_2$ standard and samples collected after 200 and 300 soap-film formation cycles. No detectable $H_2O_2$ resonance was observed in either soap-film sample; the inset shows an expanded view of the $H_2O_2$ chemical-shift region of soap film samples. **(b)** Comparison of the 1 μM $H_2O_2$ standard with the sample collected after 300 soap-film cycles and subsequently spiked with 1 μM $H_2O_2$. The clear recovery of the $H_2O_2$ resonance in the spiked soap-film matrix confirms that $H_2O_2$ remains detectable in the presence of the soap components and supports the absence of detectable $H_2O_2$ generation during soap-film formation.

Although [1]H NMR provides a direct and highly reliable approach for the molecular detection of $H_2O_2$, we examined the soap-film samples using several independent $H_2O_2$ detection methods out of curiosity. These complementary measurements were performed also to interrogate whether any of the commonly employed peroxide assays produced a positive response that could potentially be misinterpreted as $H_2O_2$ formation. Accordingly, the samples were analyzed using Amplex Red, the Hydrogen Peroxide Assay Kit (HPAK), iodometric analysis (also used in original report[83]), and potassium titanium oxalate (PTO).

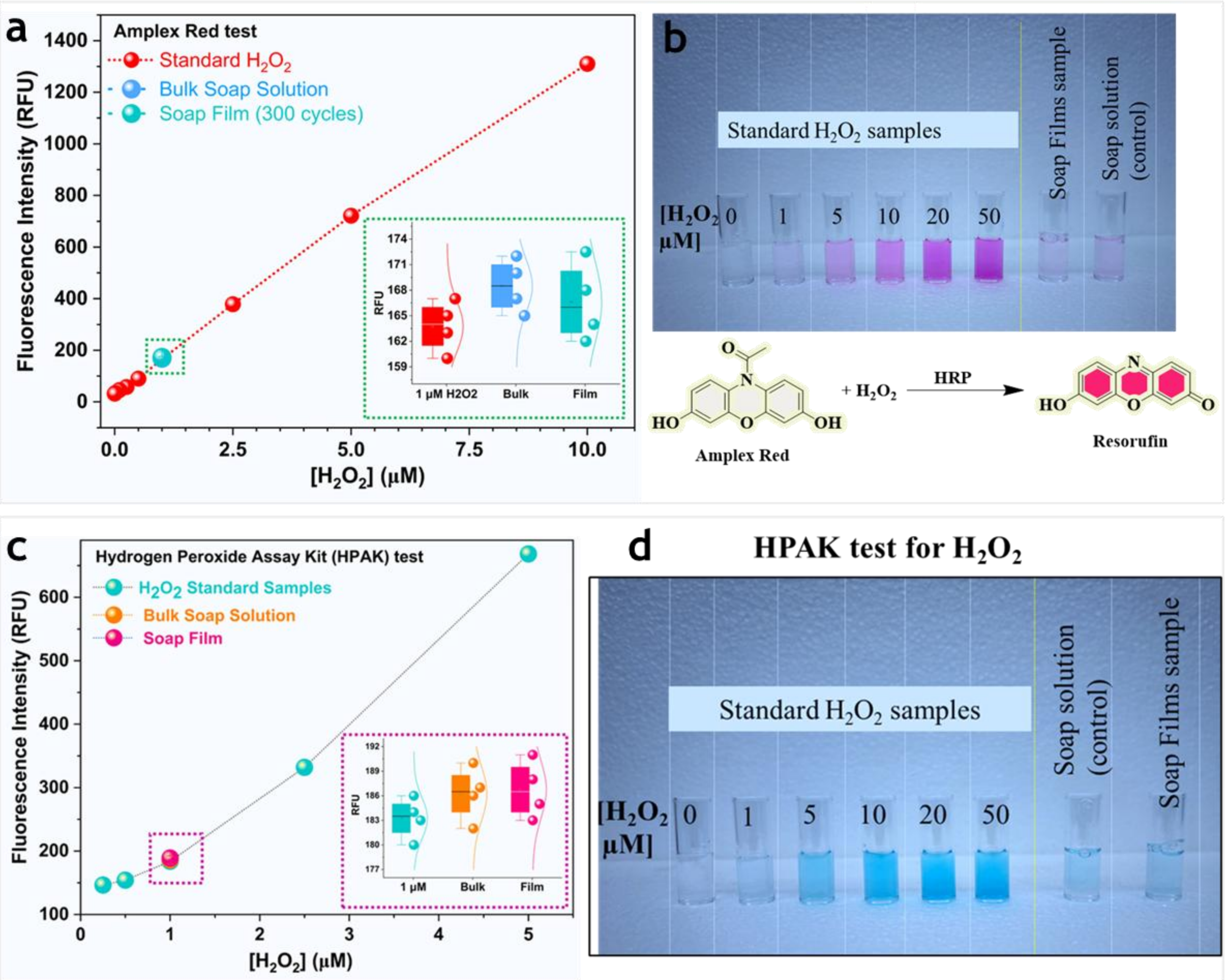

**Figure 3.** $H_2O_2$ analysis of bulk and soap-film samples using fluorescence-based assays. (a,b) Amplex Red calibration and representative responses for $H_2O_2$ standards, bulk soap solution, and the sample collected after 300 soap-film cycles. The inset in (a) shows replicate fluorescence measurements for the 1 µM $H_2O_2$ standard, bulk soap solution, and film-cycled sample. (c,d) Corresponding HPAK calibration and assay responses, with replicate measurements shown in the inset of (c). Bulk and film-cycled samples exhibited statistically comparable fluorescence responses in both assays, providing no evidence for additional $H_2O_2$ formation upon repeated soap-film generation.

The results are presented in Figures 3, S2-S3. These complementary methods further supported the absence of detectable $H_2O_2$ formation or accumulation upon soap-film generation. Interestingly, both HPAK and Amplex Red produced an apparent $H_2O_2$-equivalent response of approximately 1 µM in both the bulk soap solution and the sample collected after 300 soap-film formation cycles. Replicate measurements (Figure 3a,c, insets) showed no statistically significant difference between the bulk and film-cycled samples for either Amplex Red ($p = 0.53$) or HPAK ($p = 0.84$; two-sided Welch's t-test, $n = 4$), demonstrating that repeated soap-film formation did not measurably increase the apparent $H_2O_2$ response. Thus, repeated soap-film formation did not produce a measurable increase in the apparent $H_2O_2$ signal relative to the corresponding bulk soap solution. We therefore investigated the origin of this background fluorescence and found that it arose predominantly from Triton X-100 in water, with a smaller contribution from PBS (Figure S4). Thus, the apparent signal detected by these fluorescence-based assays reflects matrix-derived background rather than $H_2O_2$ generated during soap-film formation.

Taken together, multiple independent $H_2O_2$ detection methods converge on the observation that no measurable $H_2O_2$ is generated after repeated soap film-formation cycles, following the approach reported by Dick & co-workers[83]. These findings extend our previous observations that increasing the air–water interfacial area does not, by itself, induce spontaneous water oxidation or $H_2O_2$ formation in water microdroplets[39], salty microdroplets[40] and, most recently, microbubbles[75]. In other words, the air–water interface of soap-films is chemically stable and does not lend itself to spontaneous peroxide generation.

**Spontaneous reduction of $Au^{3+}$ ions to Au(0) nanoparticles by soap films.**

We next examined whether soap-film drive reduction of $Au^{3+}$ ions into gold nanoparticles following the experimental protocols reported by Dick & co-workers[84]. Behavior of $Au^{3+}$ ions (100 µM) in film-cycled samples (300 soap film formation cycles) were compared with that in corresponding bulk solutions, including both the bulk soap (Triton X-100 with 100 µM $HAuCl_4$) solution and 100 µM $HAuCl_4$ in pure water. This comparison allowed us to determine whether soap-film formation provides any additional driving force for Au-ion reduction beyond the spontaneous redox chemistry already occurring in the bulk solution[86]. We employed TEM and dynamic light scattering to examine the size distribution of nanoparticles, and ICP-OES for quantifying or tracking gold ions concentrations in solution phase (see Methods for sample preparation and operational details). TEM imaging revealed nanoparticle formation in both the bulk soap solution and the soap-film-collected sample (Figure 4). Consistent with the microscopy, time-resolved DLS measurements showed particle populations in both conditions over the 1–48 h observation period (Figure 4). The bulk soap solution exhibited distributions extending approximately from 10 to 80 nm, whereas the sample in which the soap films were formed and collected (300 film formation cycles) showed a broadly similar population concentrated primarily below approximately 65 nm. Although the distributions are not identical, no distinct particle population emerged exclusively upon soap-film formation. These observations show that the precursor soap solution itself supports Au nanoparticle formation and that the presence of Au nanoparticles in a collected film sample, by itself, is therefore insufficient to establish a film-specific reduction pathway.

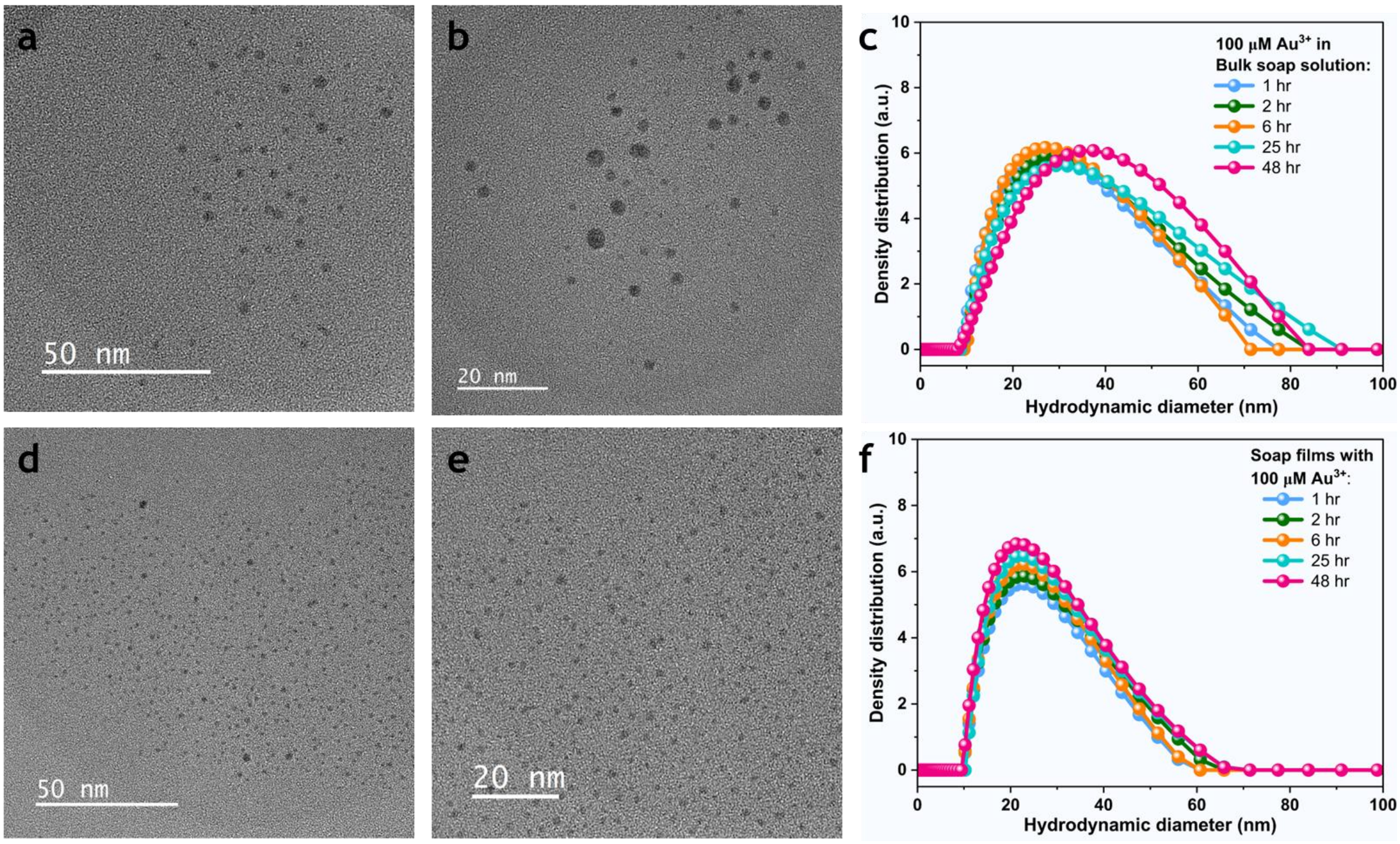

**Figure 4.** Au nanoparticle formation in bulk soap solution and following soap-film formation. (a,b) Representative TEM images of Au nanoparticles formed in bulk soap solution containing 100 μM $Au^{3+}$. (c) Time-dependent particle-size distributions measured for the bulk soap solution at 1, 2, 6, 25, and 48 h. (d,e) Representative TEM images of Au nanoparticles observed in samples collected following soap-film formation from a solution containing 100 μM $Au^{3+}$ ions. (f) Corresponding time-dependent particle-size distributions of the soap-film-collected samples. Au nanoparticles are observed under both bulk and soap-film conditions, with broadly comparable particle-size distributions, indicating that Au nanoparticle formation does not require soap-film formation.

To determine whether film formation increases the extent of Au-ion reduction, we quantified the Au remaining in solution by ICP-OES. A 100 μM $HAuCl_4$ solution corresponds to an initial elemental Au concentration of 19.7 ppm. After 1 h, the residual Au concentration in the bulk soap solution was approximately 16 ± 0.3 ppm. Critically, a comparable residual concentration was measured after 300 soap-film formation cycles (Figure 5). Thus, despite substantial loss of soluble Au (~18%) relative to the starting concentration (100 μM or 19.7 ppm), repeated formation of the soap-film interface produced no statistically significant additional Au ions loss relative to the corresponding bulk soap solution. Notably, after 1 h, the residual Au concentrations were 16.14 ± 0.32 ppm in the bulk soap solution and 16.18 ± 0.28 ppm following 300 soap-film formation cycles (mean ± SD, n = 4). No statistically significant difference was observed between the two conditions (two-sided Welch's t-test, p-value = 0.87). In conjunction with TEM and DLS, the ICP-OES data demonstrate that Au nanoparticle formation proceeds in the bulk soap solution and is not measurably enhanced by repeated soap-film formation under these conditions.

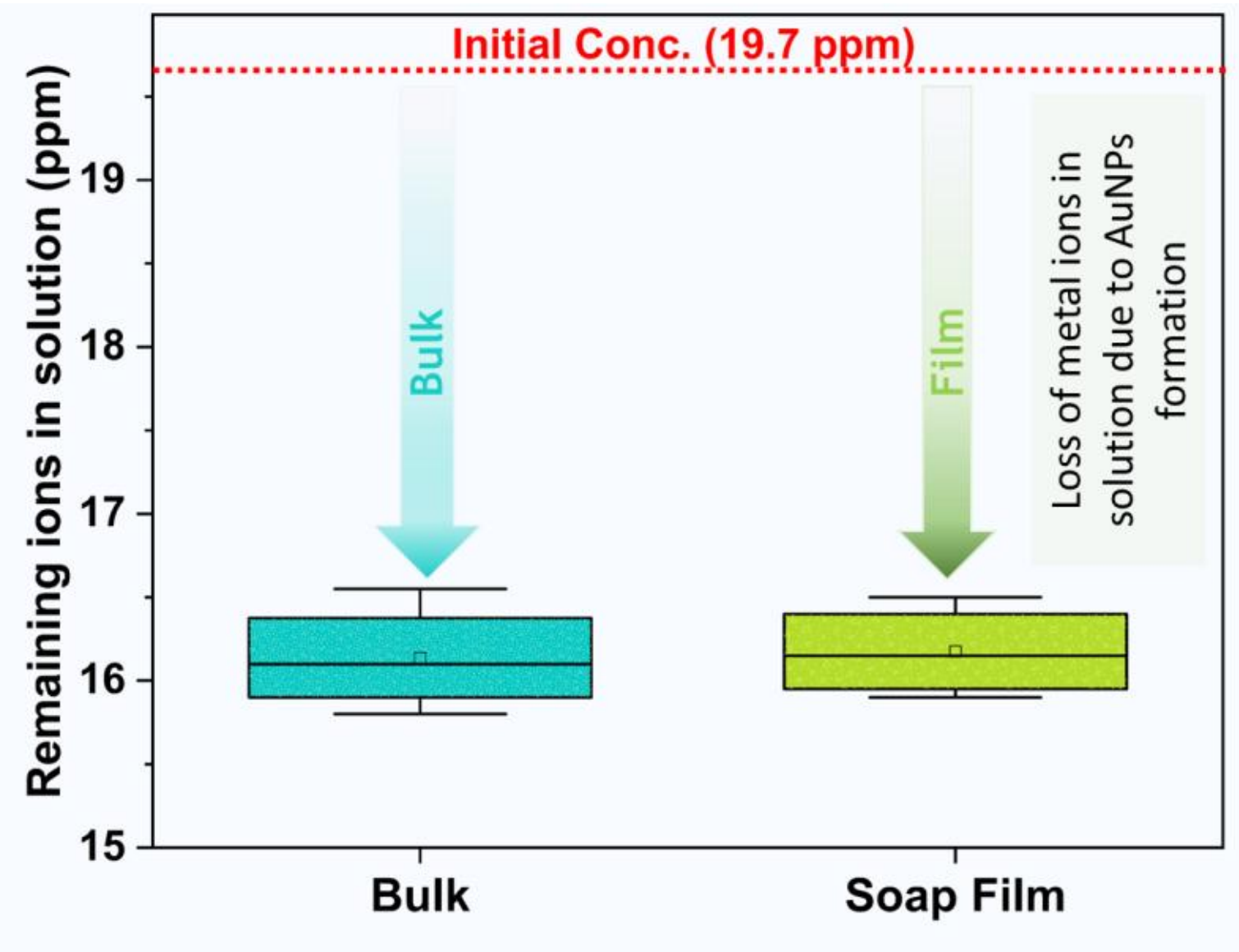

**Figure 5.** Soap-film formation does not enhance Au-ion reduction. Concentrations of Au ions remaining in solution were quantified by ICP-OES for the bulk soap solution and following 300 soap-film formation cycles. The initial Au concentration was 19.7 ppm (100 µM $Au^{3+}$). Comparable residual Au concentrations were measured in the bulk and soap-film-cycled solutions (300 film cycles), indicating that repeated soap-film formation does not measurably enhance Au-ion loss relative to the corresponding bulk solution. The decrease in soluble Au concentration is consistent with the spontaneous formation of Au nanoparticles observed independently by TEM.

We further examined whether the addition of $Au^{3+}$ or repeated soap-film formation induced detectable chemical changes in Triton X-100. Mass spectrometric analysis revealed broadly comparable spectral features for the blank soap solution, $Au^{3+}$-containing bulk solution, and corresponding film-cycled sample (after 300 film formation cycles), with no distinct species emerging upon either $Au^{3+}$ addition or soap-film formation (Figure S5). Within the sensitivity and mass range of these measurements, these results provide no evidence for substantial chemical transformation or oxidation of Triton X-100 under either condition.

To further examine the relationship between $Au^{3+}$ reduction and $H_2O_2$ formation, we systematically varied the $Au^{3+}$ concentration in the bulk soap solution from 50 to 200 µM. The amount of $H_2O_2$ formed increased progressively with increasing $Au^{3+}$ concentration, from approximately 10 µM at 50 µM $Au^{3+}$ to nearly 40 µM at 200 µM $Au^{3+}$ (Figure S6a). This concentration-dependent behavior establishes a clear association between the amount of $Au^{3+}$ present in solution and $H_2O_2$ generation, consistent with $H_2O_2$ formation being coupled to the spontaneous redox chemistry of $Au^{3+}$ ions rather than arising independently from the soap-film interface. We next directly compared $H_2O_2$ formation from 100 µM $Au^{3+}$ in the bulk soap solution with that obtained after repeated soap-film formation under otherwise identical conditions. Both systems produced approximately 20 µM $H_2O_2$, with closely overlapping replicate distributions (Figure S6b). Thus, despite repeated generation of a large air–water interfacial area, soap-film formation produced no measurable enhancement in $H_2O_2$ generation relative to the corresponding bulk solution. Together with the comparable Au-ion loss observed by ICP-OES, these results indicate that both $Au^{3+}$ reduction and the accompanying $H_2O_2$ formation are already operative in the bulk soap solution and do not require amplification of the air–water interface.

The observation that Au nanoparticles form spontaneously in bulk soap solution without soap-film generation is consistent with our previous finding that dilute $Au^{3+}$ salts can undergo spontaneous reduction in bulk water.[86] We therefore examined bulk 100 µM $HAuCl_4$ in Milli-Q water in the complete absence of Triton X-100, PBS, and deliberate film formation. TEM, HRTEM, and electron diffraction confirmed the formation of crystalline Au nanoparticles, while time-resolved DLS showed the development and growth/aggregation of particle populations in the bulk aqueous phase (Figure 6). These measurements establish that an amplified air–water interface is not a prerequisite for spontaneous Au nanoparticle formation in this system.

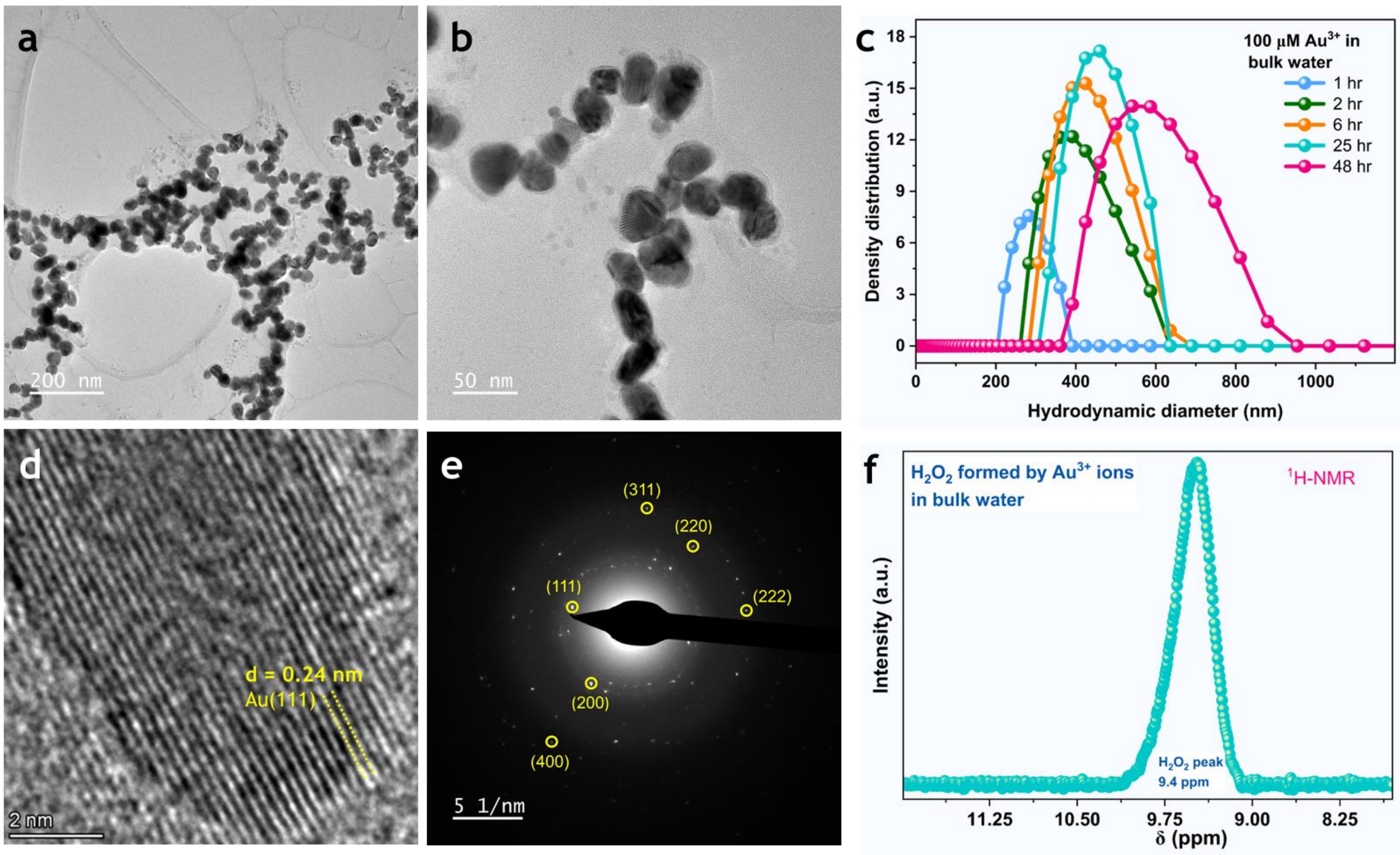

**Figure 6.** Spontaneous Au nanoparticle formation and corresponding $H_2O_2$ generation in bulk water. (a,b) Representative TEM images of particles formed following dilution of $Au^{3+}$ to 100 µM in Milli-Q water, showing the spontaneous formation of Au nanoparticles in the absence of soap-film formation. (c) Time-dependent DLS particle-size distributions of the Au-containing bulk aqueous solution measured at 1, 2, 6, 25, and 48 h; Milli-Q water is included as a blank. (d) HRTEM image showing lattice fringes with an interplanar spacing of approximately 0.24 nm, consistent with the Au(111) plane. (e) Selected-area electron diffraction (SAED) pattern showing reflections consistent with crystalline fcc Au. (f) $^1H$ NMR spectrum of the $Au^{3+}$-containing bulk aqueous solution showing the formation of $H_2O_2$.

We close by noting our $^1H$ NMR analysis of the bulk $Au^{3+}$ solution revealed concomitant $H_2O_2$ formation during Au reduction (Figure 6f). This observation provides a direct link between $Au^{3+}$-ion reduction and an accompanying oxidation process in the bulk aqueous phase. Together with our earlier work showing spontaneous reduction of $Au^{3+}$ and other sufficiently oxidizing metal ions like $Pt^{4+}$ and $Fe^{3+}$ in bulk solvents such as water, methanol & ethanol,[86] these results support a conventional coupled-redox origin for Au nanoparticle formation rather than a requirement for intrinsic soap-film reactivity. The combined controls establish a useful mechanistic asymmetry: soap-film formation in the absence of $Au^{3+}$ produces no detectable $H_2O_2$, whereas $Au^{3+}$ in bulk water produces both Au nanoparticles and $H_2O_2$ without deliberate amplification of the air–water interface. Thus, neither Au nanoparticle formation nor $H_2O_2$ detection can, on its own, be taken as evidence for soap-film-induced redox chemistry.

## Conclusion

In summary, our results demonstrate that amplification of the air–water interfacial area through soap-film formation is neither necessary for the observed $Au^{3+}$ ion reduction nor sufficient, by itself, to induce detectable $H_2O_2$ generation. Specifically, using six independent analytical methods – including the peroxide test strips and iodide-based assay employed in the original report – anchored by $^1$H-NMR with a 50 nM detection limit, we found no $H_2O_2$ attributable to soap-film formation. Importantly, spike-recovery controls confirmed that $H_2O_2$ remains readily detectable in the soap matrix, ruling out assay suppression, and the μM-level "$H_2O_2$-equivalent" responses of common fluorescence assays were traced to background contributions from the surfactant and buffer themselves. Furthermore, TEM, time-resolved DLS, and ICP-OES demonstrate that Au nanoparticles form spontaneously in the bulk soap solution – and even in plain water, accompanied by $H_2O_2$ from the coupled oxidation of water – while ~300 film cycles alter neither the particle-size distributions nor the residual $Au^{3+}$ concentrations relative to bulk controls. Together, these findings demonstrate that soap films amplify the air–water interface but neither produce detectable $H_2O_2$ nor enhance the reduction of gold ions.

## Experimental Methods

## Chemicals

All chemical reagents were used as received without further purification unless otherwise specified. Potassium dihydrogen phosphate ($KH_2PO_4$ –monobasic)(CAS-7778-77-0) , dipotassium phosphate ($K_2HPO_4$ – dibasic)(CAS - 7758-11-4), 30% hydrogen peroxide (CAS-7722-84-1) , buffer solution (pH = 7.4, potassium biphthalate buffer), hydrochloric acid (36% w/w aqueous solution)(CAS-7647-01-0), and Potassium Iodide (KI) (CAS-7681-11-0), potassium chloride (KCl) (CAS- 7447-40-7) and Sodium hydroxide (CAS-1310-73-2), were obtained from Thermo Scientific. Triton X-100 (CAS-9002-93-1), Gold chloride trihydrate ($HAuCl_4.3H_2O$) (CAS-16961-25-4) were purchased from Sigma–Aldrich. Millipore water (Millipore Milli-Q, 18.20 MΩ.cm) was used for aqueous solutions preparation. Commercial plastic wands (1-10 cm diameter) were purchased from Amazon (Brand name, ELECDON). $D_2O$ (CAS- 7789-20-0) was used as an NMR Solvent. The Amplex Red Hydrogen Peroxide/Peroxidase Assay Kit (Catalog #A22188) was purchased from Thermo Fisher and the Hydrogen Peroxide Assay Kit (HPAK, Fluorometric-near infrared) (Catalog # ab138886) was purchased from abcam.

**Soap Film Preparation**

Soap films were prepared following the experimental conditions reported by Dick and co-workers.[83] The soap solution consisted of 0.1% (v/v) Triton X-100 in 200 mM phosphate-buffered saline (PBS), with the pH adjusted to approximately 6.4 using HCl or NaOH. Experiments were also performed using commercially available 10× PBS (pH 7.4) as the buffer medium. Soap films were generated using a commercially available plastic wand, consistent with the approach employed in the previous report. To investigate whether repeated amplification of the air–water interfacial area resulted in the accumulation of reaction products, individual soap films were sequentially generated and returned to/collected into the same parent solution. This film-formation and collection procedure was repeated for up to 300 cycles.

**Peroxide Test Strips for Semi-Quantitative $H_2O_2$ Analysis**

For rapid, semi-quantitative detection of $H_2O_2$ in aqueous samples, peroxide test strips (Baker Test Strips, VWR International) with a stated detection limit of 1 ppm were employed. The strips rely on a colorimetric reaction in which exposure to $H_2O_2$ produces a blue coloration, with the intensity of the developed color providing a semi-quantitative indication of the $H_2O_2$ concentration.

**NMR Spectroscopy for $H_2O_2$(aq) Detection and Quantification**

All $^1$H NMR measurements were performed at 275 K using a Bruker Avance Neo 950 MHz NMR spectrometer equipped with a 5 mm z-gradient TCI cryogenic probe. The $H_2O_2$ proton resonance was selectively excited using a 6 ms Gaussian 90° pulse. Each scan was acquired over 53 ms, with a recycle delay of 1 ms between consecutive scans. $H_2O_2$ was identified based on its characteristic chemical shift and quantified by comparing the integrated signal intensity with those of freshly prepared $H_2O_2$ standards acquired under identical experimental conditions. A total of 50,000 scans were accumulated for each sample to maximize the signal-to-noise ratio. Under these optimized acquisition conditions, $H_2O_2$ concentrations as low as approximately 50 nM could be detected in aqueous samples (see Figure S8 for calibration curve.

**Quantification of $H_2O_2$ Using the Hydrogen Peroxide Assay Kit (HPAK)**

$H_2O_2$ concentrations in the aqueous samples were quantified using a Hydrogen Peroxide Assay Kit (HPAK) (Catalog # ab138886). The assay is based on the reaction of $H_2O_2$ with the AbIR peroxidase indicator, producing a fluorescent signal with excitation and emission maxima at 647 and 674 nm, respectively. For each measurement, 50 μL of the sample was mixed with 50 μL of the HPAK reaction mixture in a black 96-well microplate with a transparent bottom. Fluorescence measurements were performed using a SpectraMax M3 microplate reader, and the resulting data were acquired and analyzed using SoftMax Pro 7 software. $H_2O_2$ concentrations were determined from calibration curves generated using freshly prepared $H_2O_2$ standards analyzed on the same day and under identical experimental conditions as the samples.

**Hydrogen Peroxide Detection Using Amplex Red Assay**

$H_2O_2$ was quantified using the Amplex Red assay purchased from Thermo Fisher (Catalog #A22188). The assay is based on the horseradish peroxidase (HRP)-catalyzed reaction between Amplex Red reagent (10-acetyl-3,7-dihydroxyphenoxazine) and $H_2O_2$, yielding resorufin as the fluorescent product in a 1:1 stoichiometric ratio with $H_2O_2$. For each measurement, a freshly prepared working solution containing Amplex Red reagent and HRP in assay buffer was mixed with the sample and incubated in the dark at room temperature. A mixture of 50 μL of Amplex Red working solution and 50 μL of the sample was prepared in a black 96-well microtiter plate with a transparent bottom. Fluorescence was recorded using a SpectraMax M3 microplate reader with excitation and emission wavelengths of approximately 571 and 585 nm, respectively. SoftMax Pro 7 was used to process the fluorescence data. The $H_2O_2$ concentration in the samples was quantified using a calibration curve prepared from $H_2O_2$ standards and measured on the same day under identical experimental conditions.

**Particle Size Measurement**

The hydrodynamic size of the gold (Au) particles was measured by dynamic light scattering (DLS) using a Litesizer DLS 701 particle size analyzer (Anton Paar). Gold was used as the particle material with a refractive index of 0.648 and an absorption index of 3.2, while MQ water was used as the dispersant with a refractive index of 1.332. Measurements were performed for Au dispersions prepared in three different media: MQ water, soap solution, and soap + film solution. Prior to the initial measurement, each sample was shaken for 1 min to ensure homogeneous distribution of the Au particles. The sample was then transferred into a disposable cuvette and immediately placed in the Litesizer DLS 701. A 60 s equilibration time was allowed before each DLS measurement. The initial measurement was recorded at 1 h. Following the initial measurement, the cuvette containing the same sample was kept undisturbed, with no further shaking, mixing, or transfer of the solution. Subsequent particle size measurements were performed at 2, 6, 25, and 48 h to monitor changes in particle size over time. The measurements were performed at a scattering angle of 175° and a temperature of 23 °C. Three consecutive readings were obtained at each time point, and the average value was calculated and used for subsequent analysis. The experimental and instrumental conditions were maintained consistently for all samples to facilitate comparison among the different media and time points. The hydrodynamic particle size obtained from the DLS measurements was used to assess the stability of the Au particle suspensions and to determine changes in particle size over time in the different media.

**Supporting Information**

The supporting information file includes **Figure S1.** Semi-quantitative analysis of $H_2O_2$ in soap-film samples using peroxide test strips. Representative responses of peroxide test strips to freshly prepared $H_2O_2$ standards, the sample collected following repeated soap-film formation, and the corresponding bulk soap-solution control. **Figure S2,** which presents statistical comparison of fluorescence responses from bulk and soap-film samples. Replicate fluorescence measurements ($n = 4$) obtained using (a) Amplex Red and (b) HPAK for a 1 μM $H_2O_2$ standard, bulk soap solution, and the sample collected after 300 soap-film formation cycles. No statistically significant difference was observed between the bulk and film-cycled samples for either Amplex Red ($p = 0.53$) or HPAK ($p = 0.84$). **Figure S3**, which presents colorimetric detection of $H_2O_2$ in soap-film samples. (a) PTO and (b) iodometric analysis of $H_2O_2$ standards, soap-film samples, and corresponding bulk soap-solution controls. **Figure S4,** which presents the background fluorescence of the components used in the soap-film formulation. **Figure S5** presents the mass spectrometric analysis of $Au^{3+}$-containing bulk and soap-film solutions. **Figure S6** presents $H_2O_2$ formation associated with $Au^{3+}$ reduction in bulk and soap-film solutions and **Section S7** which provides the estimation of Triton X-100 concentration relative to the critical micelle concentration. In addition, **Figure S8. Presents** calibration curve and sensitivity of $^{1}$H NMR spectroscopy for $H_2O_2$ detection in pure water.

**Acknowledgments**

The authors thank Dr. Valentina-Elena Musteata from KAUST core labs for support with TEM imaging.

**Competing interests:** The authors declare that they have no competing interests.

**Data and materials availability:** All data needed to evaluate the conclusions in the paper are presented in the paper or the Supplementary Materials.

**Funding:** H.M. acknowledges KAUST for funding (Grant No. BAS/1/1070-01-01).

# Supporting Information

## Soap Films Amplify the Air–Water Interface but Neither Produce Detectable $H_2O_2$ Nor Enhance Reduction of Gold Ions

Muzzamil Ahmad Eatoo[1,2,*], Yash K. Mohature[1,2], Francisco P. Tome[1,2], Lisa Exposito[1,2], Sirisha Kamireddy[1,2], and Himanshu Mishra[1,2,*]

[1] Environmental Science and Engineering (EnSE) Program, Biological and Environmental Science and Engineering (BESE) Division, King Abdullah University of Science and Technology (KAUST), 23955-6900, Thuwal, Kingdom of Saudi Arabia

[2] Water Desalination and Reuse Platform (WDRP), Biological and Environmental Science and Engineering (BESE) Division, King Abdullah University of Science and Technology (KAUST), 23955-6900, Thuwal, Kingdom of Saudi Arabia

*Correspondence: muzzamil.eatoo@kaust.edu.sa and himanshu.mishra@kaust.edu.sa

### Section S1: Semi-Quantitative Detection of $H_2O_2$ Using Peroxide Test Strips

Peroxide test strips were employed as an initial semi-quantitative method to examine $H_2O_2$ formation during soap-film generation and to enable direct comparison with the previously reported soap-film experiments, in which peroxide-sensitive test strips were used as evidence for $H_2O_2$ formation.[1] The response of the strips was first evaluated using freshly prepared $H_2O_2$ standards spanning 0–1000 µM (Figure S1). A concentration-dependent color change was readily observed for the $H_2O_2$ standards, with increasingly intense coloration at higher $H_2O_2$ concentrations, confirming the responsiveness of the strips under our experimental conditions.

The same test was subsequently applied to the solution collected following repeated soap-film formation and to the corresponding bulk soap solution that was not subjected to film formation. In contrast to the $H_2O_2$ standards, the soap-film sample showed no discernible peroxide-dependent color development relative to the bulk soap-solution control. Thus, despite repeated generation of the air–water interface, the peroxide test strips provided no evidence for detectable accumulation of $H_2O_2$ in the soap-film sample. These observations are consistent with the results obtained using direct $^1$H NMR spectroscopy and the other complementary $H_2O_2$ detection methods employed in this study.

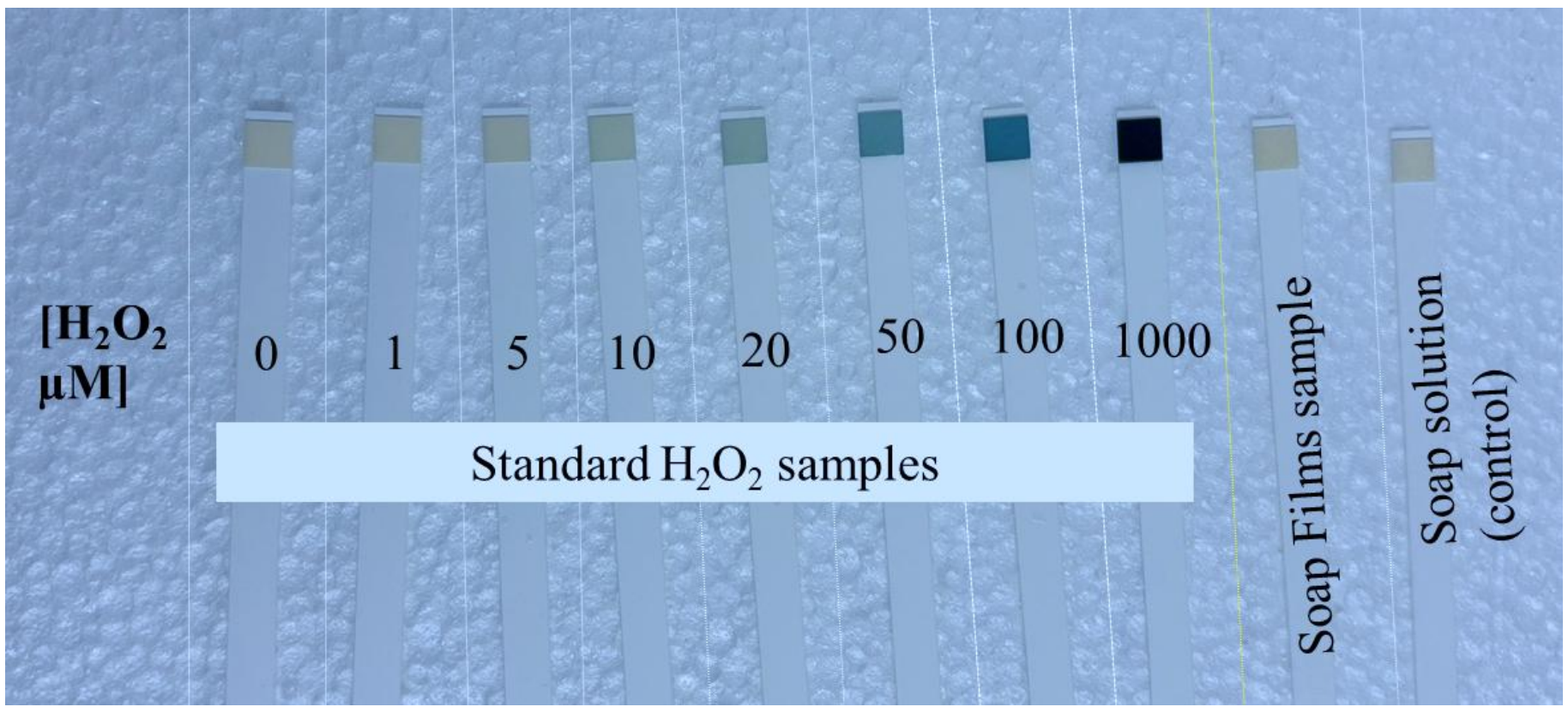

**Figure S1.** Semi-quantitative analysis of $H_2O_2$ in soap-film samples using peroxide test strips. Representative responses of peroxide test strips to freshly prepared $H_2O_2$ standards (0, 1, 5, 10, 20, 50, 100, and 1000 µM), the sample collected following repeated soap-film formation, and the corresponding bulk soap-solution control. The $H_2O_2$ standards exhibit a concentration-dependent color response, whereas no discernible peroxide-dependent color development is observed for the soap-film sample relative to the bulk soap-solution control.

**Section S2: Statistical Comparison of Fluorescence Responses in Bulk and Soap-Film Samples**

To quantitatively evaluate whether repeated soap-film formation increased the apparent $H_2O_2$ response, replicate measurements (n = 4) from the bulk soap solution and the sample collected after 300 soap-film formation cycles were compared using a two-sided Welch's t-test (Figure S2). For the Amplex Red assay, the fluorescence responses were 168.5 ± 3.1 RFU for the bulk solution and 166.6 ± 4.6 RFU for the film-cycled sample, with no statistically significant difference (p = 0.53). Similarly, HPAK measurements yielded 186.3 ± 3.3 RFU and 186.8 ± 3.5 RFU for the bulk and film-cycled samples, respectively, with no statistically significant difference (p = 0.84). Thus, both independent fluorescence assays show that repeated soap-film formation does not measurably increase the apparent $H_2O_2$ response relative to the corresponding bulk soap solution.

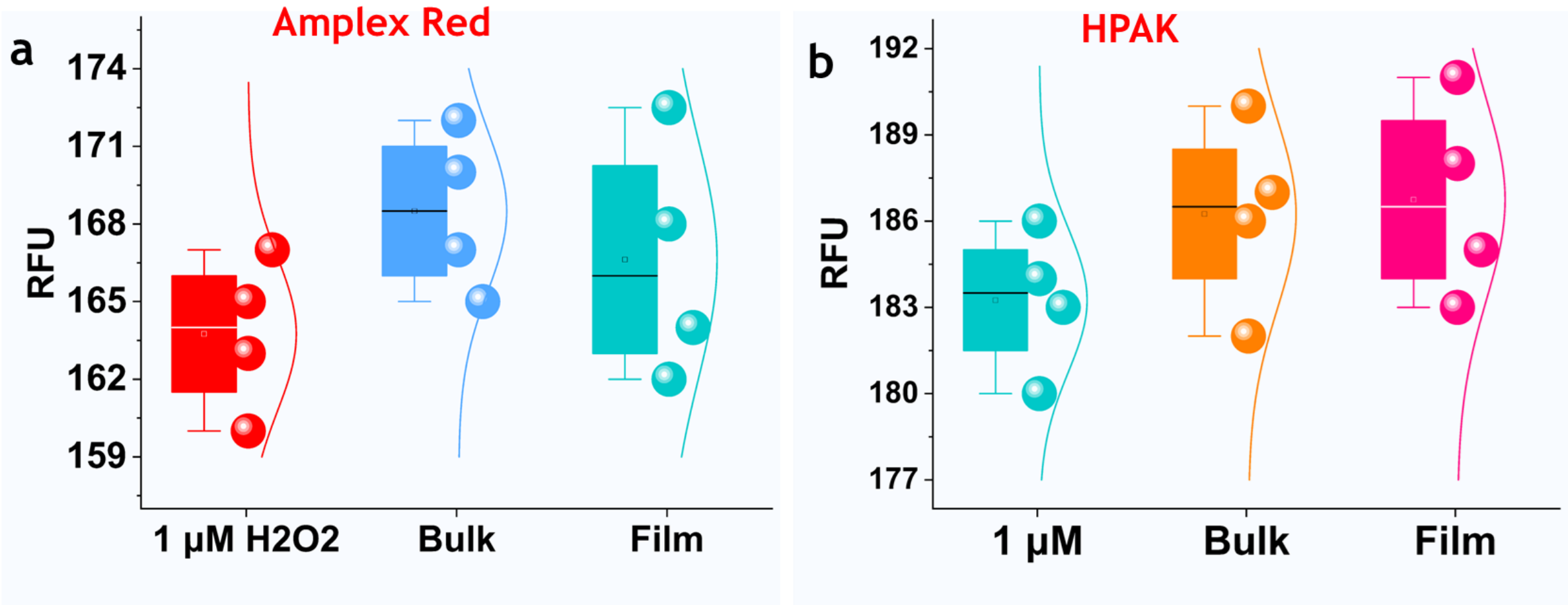

**Figure S2.** Statistical comparison of fluorescence responses from bulk and soap-film samples. Replicate fluorescence measurements ($n$ = 4) obtained using **(a)** Amplex Red and **(b)** HPAK for a 1 µM $H_2O_2$ standard, bulk soap solution, and the sample collected after 300 soap-film formation cycles. No statistically significant difference was observed between the bulk and film-cycled samples for either Amplex Red ($p$ = 0.53) or HPAK ($p$ = 0.84), as determined using a two-sided Welch's $t$-test.

## Section S3: Colorimetric Analysis of $H_2O_2$

To further verify $H_2O_2$ formation, soap-film samples and corresponding bulk soap-solution controls were analyzed using potassium titanium oxalate (PTO) and iodometric assays (Figure S3). Both methods showed concentration-dependent responses to freshly prepared $H_2O_2$ standards, whereas no discernible $H_2O_2$-dependent response was observed for either the soap-film sample or bulk control. These results further support the absence of detectable $H_2O_2$ generation during soap-film formation.

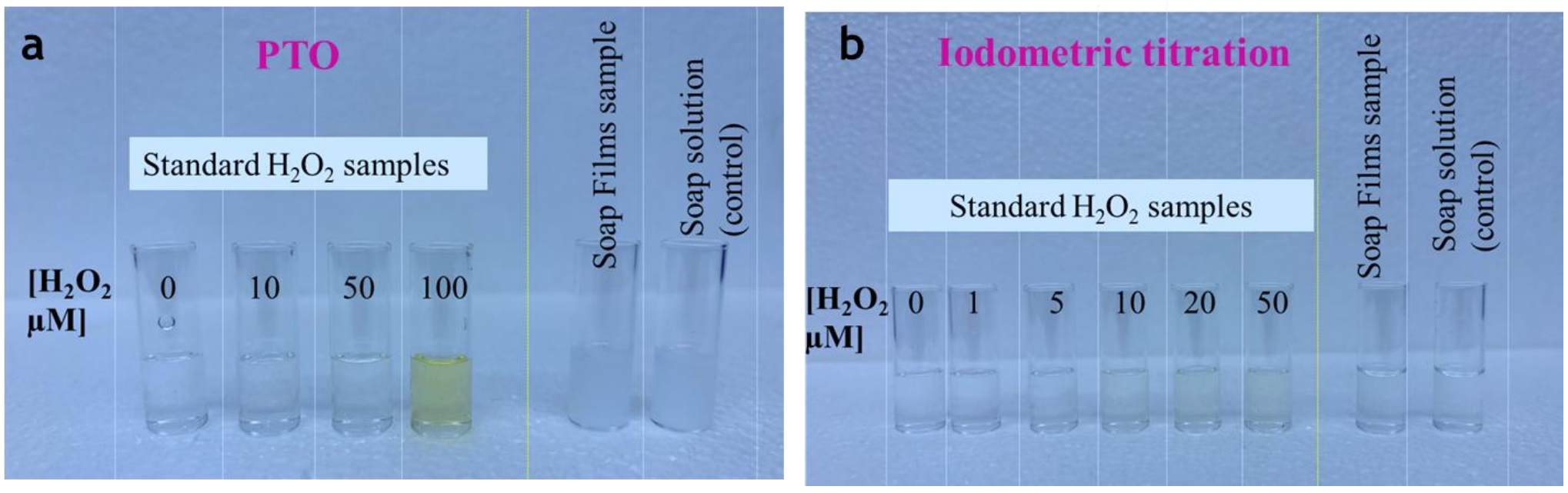

**Figure S3.** Colorimetric detection of $H_2O_2$ in soap-film samples. **(a)** PTO and **(b)** iodometric analysis of $H_2O_2$ standards, soap-film samples, and corresponding bulk soap-solution controls. No detectable $H_2O_2$-dependent response was observed in the soap-film sample relative to the bulk control.

## Section S4: Evaluation of Matrix-Derived Background in Fluorescence-Based $H_2O_2$ Detection

To determine whether components of the soap formulation contribute to the apparent fluorescence response observed during $H_2O_2$ analysis, the individual and combined matrix components were examined independently. Fluorescence intensities were measured for Milli-Q water, phosphate-buffered saline (PBS), 0.1% (v/v) Triton X-100 prepared in Milli-Q water, and 0.1% (v/v) Triton X-100 prepared in PBS (Figure S4). Milli-Q water exhibited the lowest background response, while PBS produced a modest increase in fluorescence. A substantially higher response was observed upon addition of Triton X-100 to Milli-Q water, demonstrating that the surfactant itself contributes significantly to the measured fluorescence background. The highest response was obtained for Triton X-100 in PBS, indicating an additional contribution from the PBS-containing matrix.

These control experiments demonstrate that the Triton X-100/PBS formulation used for soap-film generation possesses an appreciable intrinsic background in the fluorescence-based measurement. Therefore, low-level fluorescence responses obtained from soap-film samples cannot be assigned directly to $H_2O_2$ without comparison against the corresponding bulk soap-solution control. This matrix-derived response provides an explanation for the apparent $H_2O_2$-equivalent signal observed in the fluorescence-based assays despite the absence of a detectable $H_2O_2$ resonance by direct $^1H$ NMR spectroscopy.

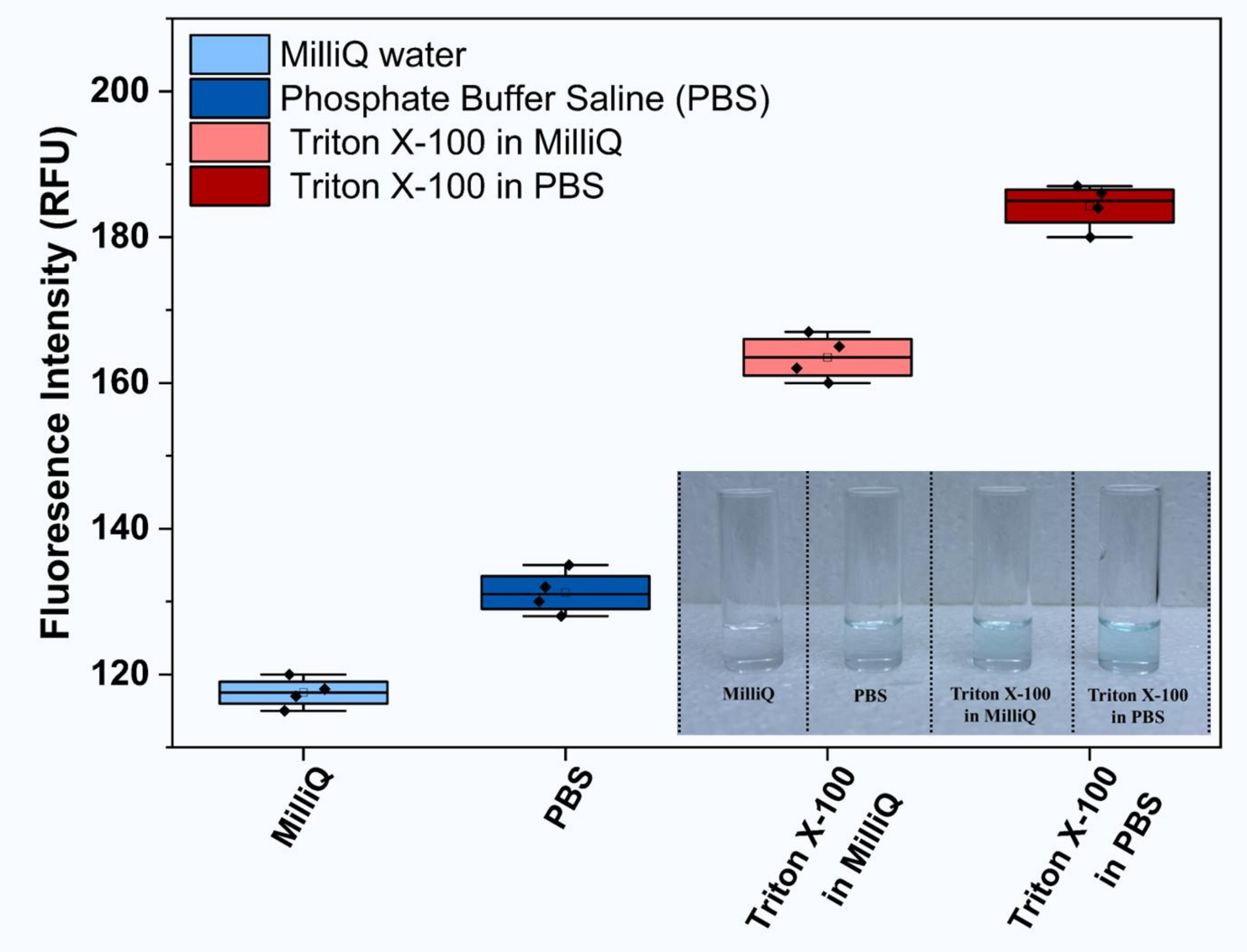

**Figure S4**. Background fluorescence of the components used in the soap-film formulation. Fluorescence intensities measured for Milli-Q water, phosphate-buffered saline (PBS), 0.1% (v/v) Triton X-100 in Milli-Q water, and 0.1% (v/v) Triton X-100 in PBS. Triton X-100-containing solutions exhibit substantially higher fluorescence than the corresponding surfactant-free controls, with the highest response observed for Triton X-100 in PBS. Individual measurements are shown as black symbols, and the inset displays representative photographs of the corresponding solutions. These results demonstrate that the Triton X-100/PBS matrix contributes an intrinsic background response that should be accounted for when interpreting fluorescence-based $H_2O_2$ assays.

**Section S5: Examination of the Chemical Stability of Triton X-100 in the Presence of $Au^{3+}$ Ions and during Soap-Film Formation**

Mass spectrometry was employed to examine whether Triton X-100 undergoes detectable chemical transformation upon addition of $Au^{3+}$ ions or during repeated soap-film formation. Mass spectra were compared for the blank bulk Triton X-100 soap solution, $Au^{3+}$-containing bulk soap solution, and the corresponding $Au^{3+}$-containing solution subjected to repeated soap-film formation (Figure S5). Across both the lower (*m/z* 150–350) and higher (*m/z* 500–820) mass ranges, the characteristic spectral features of the Triton X-100-containing solutions remained broadly comparable. Importantly, addition of $Au^{3+}$ did not produce prominent new spectral features indicative of substantial chemical modification of Triton X-100, and repeated soap-film formation did not result in additional peaks distinguishable from those already present in the corresponding bulk $Au^{3+}$/Triton X-100 solution. Within the sensitivity and mass range of these measurements, the results therefore provide no evidence for substantial chemical transformation of Triton X-100 either in the presence of $Au^{3+}$ ions or as a consequence of soap-film formation.

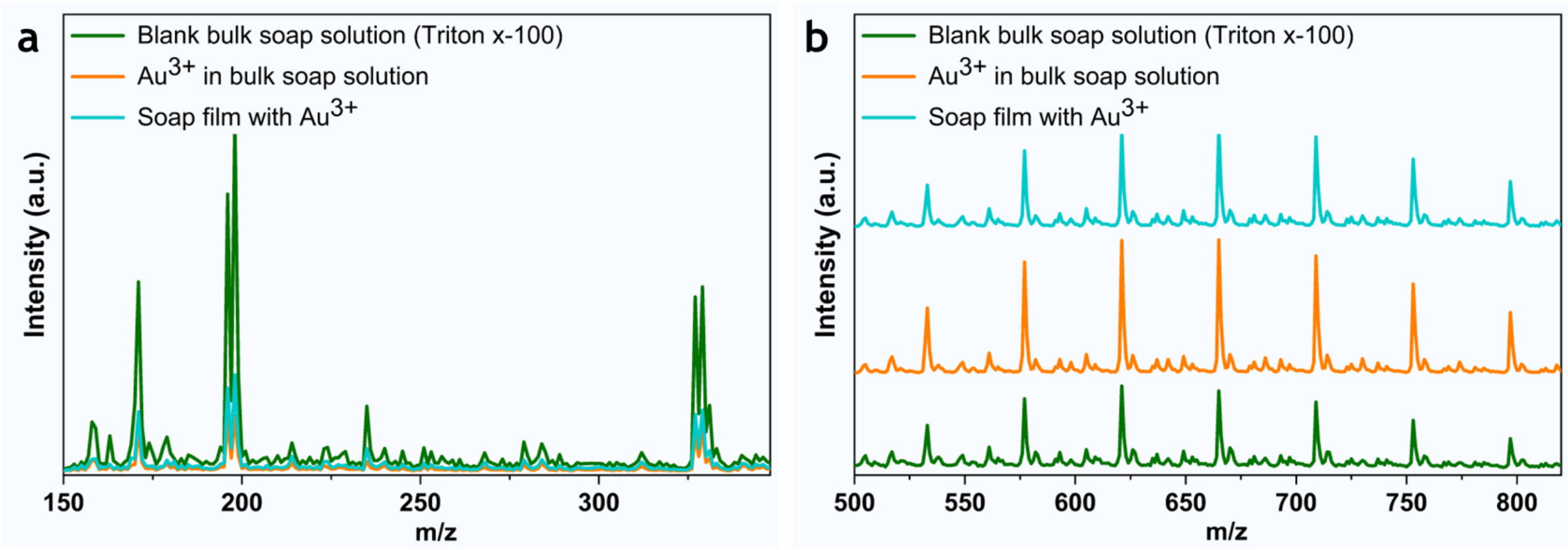

**Figure S5.** Mass spectrometric analysis of $Au^{3+}$-containing bulk and soap-film solutions. Mass spectra of the blank bulk soap solution (0.1% Triton X-100), $Au^{3+}$-containing bulk soap solution, and $Au^{3+}$-containing solution subjected to soap-film formation. Spectra are shown over the **(a)** lower *m/z* range (150–350) and **(b)** higher *m/z* range (500–820). The $Au^{3+}$ containing bulk and soap-film samples exhibit broadly comparable spectral features, while the blank Triton X-100 soap solution provides the corresponding matrix background. The similarity between the $Au^{3+}$-containing bulk and film-cycled samples provides no evidence for the appearance of distinct species uniquely associated with soap-film formation.

## Section S6: $H_2O_2$ Formation during $Au^{3+}$ Reduction

The relationship between $Au^{3+}$ concentration and $H_2O_2$ formation was examined in the bulk soap solution (Figure S6a). $H_2O_2$ formation increased systematically with increasing $Au^{3+}$ concentration from 50 to 200 µM, supporting a direct relationship between Au-ion reduction and $H_2O_2$ generation. Importantly, at 100 µM $Au^{3+}$, comparable $H_2O_2$ concentrations were measured in the bulk soap solution and following soap-film formation (Figure S6b). Thus, soap-film formation produced no measurable enhancement in $H_2O_2$ generation relative to the corresponding bulk solution.

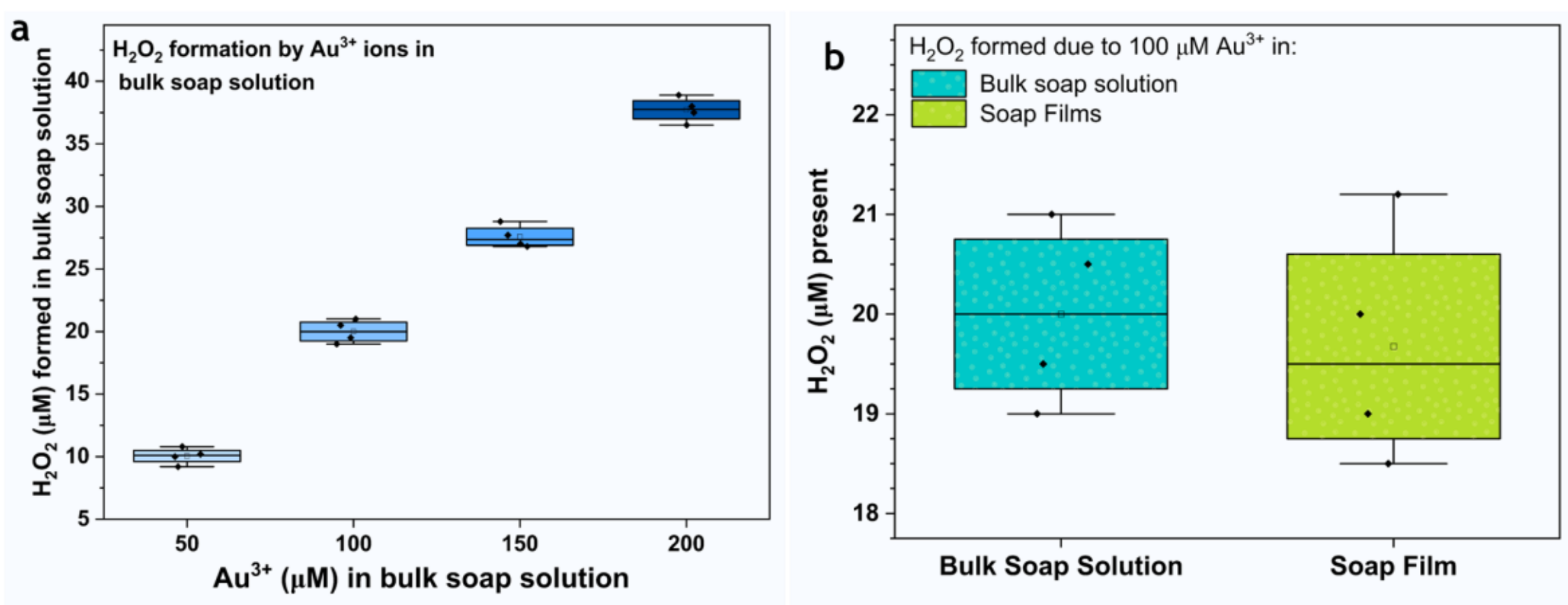

**Figure S6.** $H_2O_2$ formation associated with $Au^{3+}$ reduction in bulk and soap-film solutions. **(a)** $H_2O_2$ formation as a function of $Au^{3+}$ concentration (50–200 μM) in bulk soap solution. **(b)** Comparison of $H_2O_2$ concentrations generated by 100 μM $Au^{3+}$ in bulk soap solution and following soap-film formation. Comparable $H_2O_2$ concentrations were observed under both conditions, indicating no measurable enhancement upon soap-film formation.

**Section S7. Estimation of Triton X-100 concentration relative to the critical micelle concentration**

The concentration of Triton X-100 used to prepare the soap solution was 0.1% (v/v). To determine whether this concentration was above the critical micelle concentration (CMC), the corresponding molar concentration of Triton X-100 was estimated.

A concentration of 0.1% (v/v) corresponds to:

$0.1\% \text{ (v/v)} = 0.1 \text{ mL per } 100 \text{ mL} = 1.0 \text{ mL L}^{-1}$

Using a density of Triton X-100 of approximately 1.07 g $mL^{-1}$, the mass of Triton X-100 present per litre of solution is therefore:

$$m = (1.0 \text{ mL L}^{-1})\,(1.07 \text{ g mL}^{-1}) = 1.07 \text{ g L}^{-1}$$

Using an average molecular weight of 647 g $mol^{-1}$ for Triton X-100, the corresponding molar concentration is:

$$C = 1.65 \times 10^{-3} \text{ mol L}^{-1}$$

Thus,

$$C_{\text{Triton X-100}} \approx 1.65 \text{ mM}$$

The reported critical micelle concentration of Triton X-100 in water is approximately **0.22 mM**. Accordingly, the Triton X-100 concentration employed in the soap-film experiments was:

$$\frac{1.65 \text{ mM}}{0.22 \text{ mM}} \approx 7.5$$

or approximately 7.5 times the CMC.

Therefore, the Triton X-100 concentration used for soap-film formation was substantially above its CMC. Under these conditions, the surfactant is expected to strongly populate the air–water interface, while the excess surfactant in the bulk exists predominantly in micellar and monomeric equilibrium.

**Section S8: Calibration Curve of NMR spectroscopy for $H_2O_2$ detection**

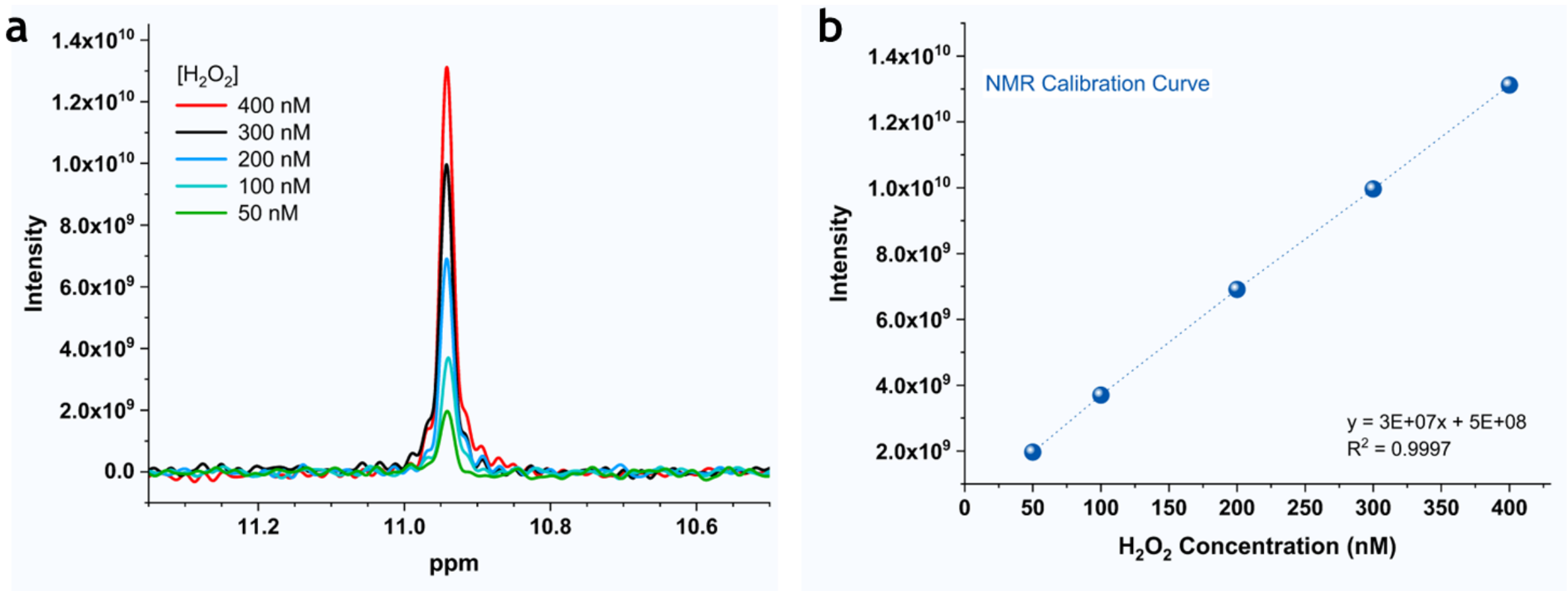

**Figure S8. Calibration and sensitivity of $^1$H NMR spectroscopy for $H_2O_2$ detection pure water.** (a) $^1$H NMR spectra of aqueous $H_2O_2$ standards at concentrations of 50, 100, 200, 300, and 400 nM, showing the concentration-dependent $H_2O_2$ resonance at approximately 10.94 ppm. (b) Calibration curve obtained from the corresponding $H_2O_2$ signal intensities as a function of concentration. The response was linear over the investigated concentration range (50–400 nM), with $y = 3 \times 10^7 x + 5 \times 10^8$ and $R^2 = 0.9997$.